\documentstyle[prl,aps,preprint,epsfig]{revtex}
\def\btabl{\begin{table}}   \def\etabl{\end{table}}
\def\bnn{\begin{eqnarray*}}   \def\enn{\end{eqnarray*}}
\def\btabu{\begin{tabular}}   \def\etabu{\end{tabular}}
\def\bec{\begin{displaymath}} \def\eec{\end{displaymath}}
\def\eqref#1{(\ref{#1})}

\newcommand{\bfig}{\begin{center}\begin{picture}}
\newcommand{\efig}[1]{\end{picture}\\{\small #1}\end{center}}

\newcommand{\beq}{\begin{equation}}
\newcommand{\eeq}{\end{equation}}
\newcommand{\nn}{\nonumber}
\newcommand{\bea}{\begin{eqnarray}}
\newcommand{\eea}{\end{eqnarray}}

\newcommand{\rfn}[1]{(\ref{#1})}
\newcommand{\Eq}[1]{Eq.~(\ref{#1})}

\newcommand{\ea}{{ et al.}}

\newcommand{\np}[1]{{ Nucl. Phys. }{\bf #1}}
\newcommand{\plet}[1]{{ Phys. Lett. }{\bf #1}}
\newcommand{\pr}[1]{{ Phys. Rev. }{\bf #1}}
\newcommand{\prlet}[1]{{ Phys. Rev. Lett. }{\bf #1}}
\newcommand{\zp}[1]{{ Z. Phys. }{\bf #1}}

\newcommand{\ptp}[1]{{ Prog. Theor. Phys. }{\bf #1}} 
\newcommand{\arnps}[1]{{ Ann. Rev. Nucl. Part. Sci. }{\bf #1}}

\newcommand{\epj}[1]{{Eur. Phys. J.}{\bf #1}}

\def\lsim{\mathrel{\vcenter{\hbox{$<$}\nointerlineskip\hbox{$\sim$}}}}
\def\gsim{\mathrel{\vcenter{\hbox{$>$}\nointerlineskip\hbox{$\sim$}}}}

\begin{document}

\preprint{\vbox{\baselineskip=13pt
\rightline{CERN-TH/98-376}
\rightline{DESY 98-208}
\rightline{FTUV/98-98}
\rightline{MZ-TH/98-58}
\rightline{hep-ph/9901265}}}

\draft

\title{CP asymmetries in $B_s$ decays and \\
spontaneous CP violation}

\author{G. Barenboim$^a$, J. Bernabeu$^b$, J.Matias$^c$ and M. Raidal$^d$
\footnote{e-mails: gabriela@thep.physik.uni-mainz.de, 
bernabeu@evalvx.ific.uv.es, \\
Joaquim.Matias@cern.ch,  raidal@mail.desy.de}} 
\vskip -0.5cm
\vspace{-0.5cm}
\address{{\small 
$^a$ Institut f\"ur Physik, Johannes Gutenberg Univertit\"at, D-55099,
Mainz, Germany \\
$^b$ Departamento de F\'{\i}sica Te\'orica, Universidad de Valencia,
E-46100, Burjasot, Valencia, Spain\\
$^c$Theory Division, CERN, CH-1211, Geneva 23, Switzerland\\
$^d$DESY, Theory Group, Notkestra{\ss}e 85, D-22603 Hamburg, Germany }}

\vskip -1cm

%\date{\today}

\maketitle 

\begin{abstract}
We study possible effects of new physics in CP asymmetries in 
two-body $B_s$ decays in left-right models with spontaneous CP violation.
Considering the contributions of new CP  phases to the $B_s$ mixing
as well as to the penguin dominated decay amplitudes we show that, 
with the present constraints, large deviations from the standard model 
predictions in CP asymmetries are allowed in both cases.
Detection of the new physics can be done by measuring non-zero asymmetries 
which are predicted to vanish in the standard model or by comparing
two measurements which are predicted to be equal in the standard model. 
In particular, we show that the measurement of the CKM angle 
$\gamma$ in electroweak penguin dominated processes 
$B_s^0\to\rho^0\eta^{(')},\,\rho^0\phi$ can  largely be affected by 
the new physics.   
\end{abstract}

\vspace{1.5cm}

\leftline{December 1998}

\newpage 

\section{Introduction}
Although the establishment of CP violation goes back to more
than thirty years ago, CP violation stills remains a mystery \cite{quinn}. 
It has 
been observed so far only in the kaon system and our entire knowledge can
be summarized by the single CP violating quantity $\epsilon_K$.
In the standard model (SM) the CP violation
can be accommodated due to the single phase of the CKM matrix \cite{ckm},
whereas extended models, in general, contain extra sources of CP violation.
Some of them explain the origin of CP violation due to 
spontaneous symmetry breaking \cite{lee}.
These models include  the two Higgs doublet models \cite{2hdm}, aspon
models \cite{aspon}, models with ``real CP violation" \cite{yanagida} and 
left-right symmetric models \cite{lr}
(LRSM). 
This possibility is particularly natural in the LRSM since parity 
is a spontaneously broken quantity in these theories. 
Because the SM has specific predictions on the size as well as on the 
pattern of CP violation in $B_{d,s}$ meson decays these predictions
can be tested in the future $ B$-factories and dedicated experiments 
HERA-b and LHC-b. 

On the experimental side there might already exist some evidence for
non-standard flavour physics in the quark sector. The large rate of
$B\to\eta'X_s$ measured by CLEO 
seems to require enormously enhanced
$b\to sg$ \cite{kagan} if compared with the SM prediction.
If  new CP phases are incorporated in order to explain  
the enhancement of $b\to sg$,
the CP asymmetries in  $B_{d,s}$ decays would turn out to 
be largely affected.
Such a situation may occur in models of supersymmetry 
\cite{effsusy},
models with enhanced chromomagnetic operators \cite{kagan2}
as well as in the LRSM \cite{meie}, motivating also the present work.

 In the SM the $B_{d,s}$ systems have  been extensively studied \cite{burf}.
 There are also a number of 
studies of the new physics  
effects in $B_d$ decays \cite{gross,susy}. However, 
the $B_s$ system has received somewhat less attention from new physics
point of view \cite{gros,bs}. 
Very fast oscillations of the $B_s$ system require
outstanding experimental sensitivity (not yet achieved) 
 to measure  time dependent asymmetries.
However, due to large width difference $\Delta \Gamma^{(s)}$ the $B_s$
system offers new possibilities for testing new physics  which do not
exist in the $B_d$ system.   
Moreover, since many CP asymmetries in $B_s$ decays
are predicted to be vanishingly small in the SM 
the new physics effects may easily show up.

In this paper we study the possible new physics effects
in CP asymmetries in two body $B_s$ decays in the LRSM with spontaneous
CP violation \cite{cpk,cpb}. In this model there are new CP phases which are
unsuppressed for the third family and thus may affect $B_s$ system.
Although the new phases of the right-handed sector can appear at tree
level, these contributions are strongly suppressed by current phenomenology.
So we look for new physics effects only in loops. We shall consider
both new physics contribution to the $B_s^0-\bar B^0_s$ mixing and to the 
penguin dominated decay modes and show that the new physics 
in both cases may be observable. 
While the former effect  is suppressed by large 
right-handed gauge boson and flavour-changing Higgs bosons masses 
 the latter is
suppressed by the left-right mixing angle. This is because 
in the decay amplitudes  the dominant new contribution  arises due to 
new dipole operators induced by penguin diagrams with $W_L$ in the loop
which interacts via $(V+A)$ currents in one vertex. Thus the dipole operators
contributions to the CP asymmetries in LRSM are enhanced by   
\begin{enumerate}
\item[$ (i)$] the large ratio $m_t(M_Z)/m_b(M_Z)$, 
\item[$ (ii)$]  the larger values of the Inami-Lim type 
loop function if compared with the SM,
\item[$ (iii)$] by {\it two independent} new phases $\sigma_{1,2}$ which
values are unconstrained.
\end{enumerate} 
The new effect is dominated by the gluonic penguins. However,
for the decays with vanishing QCD penguin contributions like 
$B_s \rightarrow \eta^{(')}\pi,\,\phi\pi,\,\eta^{(')}\rho,\,\phi\rho,$
the electromagnetic dipole operators may play an important role. 
It has been proposed \cite{fleischer2} 
to use these decay modes to measure the CKM angle 
$\gamma.$ We shall show that in some of the decays 
these measurements can be dominated by
the new phases occurring in dipole operators.

The paper is organized as follows. In the next Section we discuss the  
$B_s$ system in the presence of new physics and study LRSM
contributions to the $B_s$-$\bar B_s$ mixing. In Section 
III we study new physics contributions to the decay amplitudes. We conclude
in Section IV.

\section{CP asymmetry in the $B_{s}$ system}

The general expression of the time-dependent CP asymmetry for the decays
that were tagged as pure   $B^{0}_{q}$
or ${\bar B}^{0}_{q}$ (with $q=d,s$) into CP eigenstates,
\bea
a^{(q)}_{CP}(t)\equiv {\Gamma(B_q^0(t) \rightarrow f) -
\Gamma({\bar B_q^0}(t) \rightarrow f) \over \Gamma(B_q^0(t) \rightarrow f)
+\Gamma({\bar B_q^0}(t) \rightarrow f)}\,,
\eea
is given explicitly by 
\bea \label{asym}
a^{(q)}_{CP}(t)=-{ \left( |\lambda^{(q)}|^2-1 \right) 
{\rm cos} (\Delta M^{(q)} 
t) - 2
{\rm Im} \lambda^{(q)} {\rm sin} (\Delta M^{(q)} t)  \over 
(1+ |\lambda^{(q)}|^2) {\rm cosh} ({\frac{1}{2} \Delta \Gamma^{(q)} t}) -2 {\rm
Re}
\lambda^{(q)}
{\rm sinh} ({\frac{1}{2} \Delta
\Gamma^{(q)} t })} \,,
\eea
where $\Delta \Gamma^{(q)}=\Gamma_H^{(q)}-\Gamma_L^{(q)}$ and 
$\Delta M^{(q)}=M_H^{(q)}-M_L^{(q)}$ are the differences in  decay
rates and masses  between the physical eigenstates, respectively, 
 whereas $\lambda^{(q)}$ is given by
\bea \label{lam}
\lambda^{(q)}=\left( \sqrt{\frac{M_{12}^{(q)*}-\frac{i}{2}\Gamma_{12}^{(q)*}}
{M_{12}^{(q)}-\frac{i}{2}\Gamma_{12}^{(q)}}}\right)\frac{\bar A_{q}}{A_{q}}=
e^{-2i\phi_M^{q}}
\frac{\bar A_{q}}{A_{q}}\,.
\label{lambda}
\eea
Here
$A_{q}$ and $\bar A_{q}$ are the amplitudes of $B^0_{q}$ and $\bar
B^0_{q}$
decay to a common CP eigenstate, respectively, and we have used 
$|\Gamma_{12}^{(q)}|\ll |M_{12}^{(q)}|$ to introduce the $B^{0}_{q}$-$\bar
B^{0}_{q}$ mixing phase $\phi_M^{q}$.

When this asymmetry concerns the $B_d$ system the terms $\Delta
\Gamma^{(d)}$ in the denominator are neglected and the asymmetry takes
a simple form (see, for instance, \cite{fb} for a review) 
\bea 
a^{(d)}_{CP}(t)=-{ \left( |\lambda^{(d)}|^2-1 \right) {\rm cos}
 (\Delta M^{(d)} t) - 2
{\rm Im} \lambda^{(d)} {\rm sin} (\Delta M^{(d)} t)  \over
(1+ |\lambda^{(d)}|^2)}\,,
\eea
whereas in the $B_s$ system this is not possible since 
$\Delta \Gamma^{(s)}/\Gamma^{(s)}$ is expected  to be ${\cal O}(20 \%)$
at leading order\footnote{
The recent next-to-leading order results \cite{bbgln}, however, 
tend to reduce the central value of $\Delta \Gamma^{(s)}/\Gamma^{(s)}$.}
 in the SM \cite{width}.
This will affect the
time dependent CP asymmetry already at $t\ge 2/\Delta \Gamma^{(s)}$.

The integrated asymmetry is now, including the $\Delta \Gamma^{(q)}$ terms:
\bea \label{int}
{A}^{(q)}_{CP}={\left(-1+|\lambda^{(q)}|^2-2{\rm Im} \lambda^{(q)} x_q 
\right)(-4 
+ y_q^2)
\over 4 (1 + x_q^2) \left(1+ |\lambda^{(q)}|^2-{\rm Re} 
\lambda^{(q)} y_q \right) }\,,
\eea
where $x_q=\Delta M^{(q)}/\Gamma^{(q)}$ and $y_q=\Delta
\Gamma^{(q)}/\Gamma^{(q)}$.
Notice that the previous expression reduces to the well known integrated
asymmetry of the $B_d$ system $A^{(d)}_{CP}$ in the limit $y_d \rightarrow 0$.
Moreover if $\lambda^{(q)}$ reduces itself
 to a pure phase the contribution of the
width difference factor outs, i.e., 
$$A^{(s)}_{CP}({\rm Re} \lambda^{(s)}=0)=A^{(s)}_{CP}(
y_s \rightarrow 0) (1 -y_{s}^2/4).$$

A first measurement of $y_s$ comes out from Fermilab \cite{ferm}, 
$y_{s}=0.34^{+0.31}_{-0.34},$ 
and they conclude that with the current statistics they are not sensitive
to $B_s$ lifetime differences, but they give an upper bound
at a $95\%$  confidence level of $y_{s} < 0.83.$

It is clear  that the CP asymmetry in the $B_s$ system is sensitive
also to $\Delta \Gamma^{(s)}$. Then if some new physics affects it, 
it will also affect
the asymmetry. 
Now, if this contribution is CP violating, it leads,
as shown by Grossman in \cite{gros}, to an unavoidably  reduction 
of the width difference with respect to the SM prediction. In general,
 the width difference is given by
\bea
\Delta \Gamma^{(s)}={4 Re(M_{12}^{(s)} 
\Gamma^{(s)*}_{12} )\over \Delta M^{(s)}}\,.
\eea
The experimental lower bound $\Delta M^{(s)} > 9.5\, {\rm ps}^{-1}$  \cite{low}
implies that $\Delta M^{(s)} \gg \Delta \Gamma^{(s)} $
and consequently
$\mid M_{12}^{(s)} \mid \gg \mid \Gamma_{12}^{(s)} \mid  $.
Therefore to a very good approximation
\bea
\Delta M^{(s)}=2 |M_{12}^{(s)}|\,,
\eea
and 
\bea
\Delta \Gamma^{(s)}= 2 |\Gamma_{12}^{(s)}| 
|{\rm cos} 2 \xi| \quad ,\quad 2 \xi\equiv
arg(-M_{12}^{(s)} \Gamma^{(s)*}_{12})  \label{for1}\,.
\eea
In the SM $\cos(2 \xi) \sim 1$ whereas
any type of new physics, regardless its origin, 
will imply  $\cos(2 \xi) < 1$ reducing the value of 
$\Delta \Gamma^{(s)}$ according to \Eq{for1}. 
Although $\Delta \Gamma^{(s)}$ is not by itself an indication of CP
violation, such a reduction can come from new CP phases.
As the large SM prediction for  $\Delta \Gamma^{(s)}$
is based on the fact that the decay width into CP even final states
is larger than into CP odd final states, the appearance of new 
 phases in the mixing amplitude allows the mass eigenstates
to differ significantly from the CP eigenstates. In this way both
mass eigenstates are allowed to decay into the CP even final state
and $\Delta \Gamma^{(s)}$ reduces accordingly.

At this point it is important recall the reader that for 
$|\Gamma_{12}^{(s)}|$ to
be significatively enhanced, one needs a new decay mechanism
which dominates over the $W_1$ mediated tree decay. This is most 
unlikely, there seems to be no viable model that suggest such a situation.
Within the LRSM, the tree level exchange does not
contribute to $\Gamma_{12}^{(s)}$ and all the contributions coming from the
left-right boxes have either a $\beta=M^2_{1}/M^2_{2}$ 
  or a $\beta^2$ suppression factor when
compared to the SM result. Here $M_1$ is the mass of the normal left-handed
gauge boson and $M_{2}\gsim 1.6$ TeV \cite{beall} the mass of 
the new right-handed gauge boson.
 So, the left-right contribution is
completely negligible. We have explicitly checked that it amounts, at most -
being dreadfully optimistic-, a one per cent increase and we 
can therefore safely
take $|\Gamma_{12}^{(s)}| = |\Gamma_{12}^{(s) \, SM}|$.

The smoking gun in the left-right symmetric models comes from the
additional terms to $B_s - \overline{B_s}$ mixing \cite{cpb}. While the SM 
contribution  to the off diagonal mixing matrix element, $M_{12}^{(s)}$, is
dominated by the tt exchange, the left-right contribution 
gets its weight mainly from the $W_1 -W_2(S_2)$ boxes (being $S_2$ the
Goldstone boson which becomes the longitudinal component of $W_2$)
and the tree level flavour changing Higgs exchange 
(we denote its mass by $M_H$). The total 
 off diagonal mixing matrix element  $M_{12}^{(s)}$ can therefore be 
written as,
\bea
M_{12}^{(s)}= M_{12}^{(s)\,SM} +  
M_{12}^{(s) \,LR} = M_{12}^{(s) \,SM}\left(1+ \kappa e^{i \sigma_s} \right),
\eea
where 
\bea
\kappa  &\equiv & \left| \frac{M_{12}^{(s) \,LR}}{M_{12}^{(s) \,SM}} 
\right| 
\simeq  \left[ 0.21 + 0.13 \; \log\left(\frac{M_2}{1.6 {\mbox{TeV}}}\right)
\right] \left(\frac{1.6 {\mbox{TeV}}}{M_2}\right)^2 +
\left(\frac{12 {\mbox{TeV}}}{M_H}\right)^2. 
\eea
The new phase $\sigma_s$ can be expressed as
\bea
\sigma_s = \mbox{Arg} \left(  \frac{M_{12}^{(s)\,LR}}{M_{12}^{(s) \,SM}}
\right),
\eea 
with
\bea
\sin\sigma_{s} \simeq  \pm r \sin\alpha \left(  \mu_c/\mu_s 
 + \mu_t/\mu_b  \right).  
\label{sig}
\eea
Here $\mu_i = \pm m_i$, $r$ is the ratio of vevs occurring in the
bidoublet and $\alpha$ is the spontaneous CP violating phase.  The
approach we  follow in performing our calculations has been 
already discussed in  \cite{cpb} and we  refer the interested reader
to these papers for calculational details as well as for an explanation of
the sign factors in the masses.
It is worth pointing out that the most relevant effects we would 
find are related to the fact that despite $r$ is bounded to be smaller that
$m_b/m_t$ in order to give to the quarks their masses appropriately, the
enhancement factor $m_t/m_b$ in Eq.(\ref{sig}) ensures that
$\sigma_s$ can take any value from 0 to 2$\pi.$ Therefore the angle  
2$\xi$  given in terms of the above defined  parameters by
\bea
2 \xi={\rm ArcTan} \left( {\kappa \sin \sigma_{s}  \over 1 + \kappa 
\cos \sigma_{s}} \right)
\eea
can depart significantly from its SM value.

This can be observed in figures 1 and 2 where we plot $\cos(2\xi )$
(for the $B_s$ system) as 
a function of the spontaneous CP violating phase $\alpha$ for a fixed Higgs
boson mass of 12 TeV and for various choices of the right-handed gauge boson
mass (figure 1) and for a fixed value of the right-handed
boson mass equal to 1.6 TeV and different Higgs boson masses (figure 2).
According to the plots, the new physics can substantially affect the 
width difference for a wide range of the right-handed
particle masses.  It is worth to point
out that the Higgs and box contributions add up constructively.
The maxima observed in figures 1 and 2 correspond to the values of
$\alpha=0,\pi$ where $\sin \sigma_s$ vanishes, returning for
$\cos \xi$ to the same value as in the SM. On the contrary, the minima
in figure 1 correspond to the values of $\alpha$, $M_{W_2}$ and $M_H$ 
such that the function $1 + \kappa \cos \sigma_{s}$ vanishes. 
Comparing figure 1 and figure 2  one can see that  
the left-right contribution of the tree level Higgs exchange dominates 
over the left-right box diagrams.
The largest
possible departure from the SM value ($\cos(2 \xi) \sim 1$) is governed by the
values of the new Higgs mass 
close to its present lower bound, where $\kappa \sim 1$.

The above result, implies that within the left-right symmetric model, 
we still expect a large mixing parameter $x_s$
as within the SM.
This fact implies very rapid oscillations between $B^0_s$ and
$\overline{B^0_s}$ and therefore for keeping track of the 
$(\Delta M^{(s)}t)$ terms an outstanding experimental sensitivity 
(not yet reached) is essential. Nevertheless as was pointed out in 
\cite{burf}, this is not the end of the $B_s$ saga. 
The  untagged $B_s$ rates,
which are defined by
\bea
\Gamma(f) =
\Gamma(B_q^0 \rightarrow f)
+\Gamma({\bar B_q^0} \rightarrow f)
 \propto & \exp \left[ -\Gamma^{(s)} t
\right] \cosh \left[ \frac{\Delta \Gamma^{(s)} t}{2}
\right],
\eea
can be a method to get an insight into the mechanism of CP violation.
This possibility which does not exist in the $B_d$ case, is given 
precisely by the sizeable width difference $\Delta \Gamma^{(s)}$.

\section{New physics in decay amplitudes}

Up to this point we have seen that non-standard model CP violating 
effects could be revealed by testing whether measurements agree with
the SM allowed range. However, processes for which the SM 
contribution vanishes (or is negligibly small) offer an important
complement for these studies. In this case, any observation or 
non-observation of CP violation can be interpreted directly as a
constraint on physics beyond the SM. From this point of view
a measurement of the CP asymmetries in the decay modes 
\bea
 b \rightarrow c\overline{c}s \;\;\; && \left(\mbox{e.g.} \;\; B_s
\rightarrow \psi\phi \right),\nn\\
\label{treed}
 b \rightarrow c\overline{c}d \;\;\; && \left( \mbox{e.g.} \;\; B_s 
\rightarrow \psi  K_s \right),\\
\ b \rightarrow c\overline{u}d \;\;\; && \left( \mbox{e.g.} \; \; B_s 
\rightarrow D_{CP}^0 K_s \right),\nn
\eea
is of great interest. (It is important to notice that CP asymmetries
into final states that contain $D_{CP}$ are not going to be affected 
by the new contribution in $D -\overline{D}$ mixing).

These CP asymmetries measure the same angle of the unitarity triangle,
$\beta^\prime$ which is approximately equal to zero in the Standard
Model. In the presence of new contributions to the $B^0 - \overline{B^0}$
mixing matrix, the CP asymmetries in these modes would no longer be 
measuring the CKM angle $\beta^\prime$. However, they would all still
measure the same angle  $\beta^\prime + \delta_m$, where $\delta_m$
is the new contribution to the $B^0 - \overline{B^0}$ mixing phase. In 
the LRSM with spontaneous CP violation, as
we have already seen, $\delta_m = 2\xi$ and therefore large departures
from the expected zero are possible.
Specially interesting in this respect are the decays where the SM
contribution is already tree level and therefore are very unlikely
to be significantly affected by new physics in the decay. On the contrary,
as we have seen, the mixing amplitude can be easily modified by
new physics, providing this way an excellent testing ground
for a measurement of $\sin 2\xi $. A particularly promising example,
for future $B$-physics experiments to be performed at hadron machines,
is $B_s \rightarrow \phi \psi$ where new physics in 
the mixing can be explored.

On the other hand, new physics, in general independent from that influencing 
 $B^0 - \overline{B^0}$ mixing, 
can enter also in the decay amplitudes 
of $b$ quarks and cause deviations from the SM prediction even in the case
of vanishing new physics contribution to the  $B^0 - \overline{B^0}$ mixing. 
The decays \rfn{treed} are dominated by tree level $W$ exchange and the 
new
physics contribution to them is negligible. However, pure QCD penguin
decays like 
\bea
 b \rightarrow s\overline{s}s \;\;\; &&\left(\mbox{e.g.} \;\; B_s 
\rightarrow \phi\phi \right),\nn\\
\label{qcdd}
 b \rightarrow s\overline{d}d \;\;\; &&\left( \mbox{e.g.} \;\; B_s 
\rightarrow \bar K K_s \right),\\
\ b \rightarrow d\overline{s}s \;\;\; &&\left( \mbox{e.g.} \; \; B_s 
\rightarrow \phi K_s \right),\nn
\eea
or the electroweak penguin dominated decays
\bea
B_s \rightarrow \eta\pi,\;\eta'\pi,\;\eta\rho,\;\eta'\rho,\;\phi\pi,
\;\phi\rho\,,
\label{ewd}
\eea
may receive considerable contribution from new physics.
The SM CP asymmetries in $b\to s\bar ss$ and  $b\to s\bar dd$ decays are
vanishing while  $b\to d\bar ss$ decays should  measure the CKM angle 
$\beta.$  The decays \rfn{ewd} receive small contribution from 
tree level $W$ processes which are sensitive to the CKM angle $\gamma.$ 
It has therefore been proposed by Fleischer \cite{fleischer2} to determine
$\gamma$ from the CP asymmetries in these decays.
Since the CP asymmetries depend on the parameter $\lambda$ in \Eq{lam}
it is clear that 
the  new contributions to the $b$ quark decay amplitudes
would affect differently each of the modes 
in \Eq{qcdd},\rfn{ewd} and therefore each
of them would measure a different CP violating quantity. 
Therefore, new physics contribution to the decay amplitudes 
can be traced off not only for vanishing SM CP asymmetries but also by  
comparing the measurements of different $B_d$ and $B_s$ decay modes
which should be equal within the SM.
Complementary information on this new physics in the decay 
amplitudes can be extracted from the comparison of any of the 
penguin dominated diagrams (where the new physics in the decay 
amplitude can play an important role) with the ones only affected
by new physics in the mixing, e.g., with
$B_s \rightarrow \phi \psi$. Such a comparison would allow 
a clear separation between new physics coming from the decay and
new physics in the mixing.

The flavor changing decays $b\to q\bar q'q'$ where $q,q'=s,d$ are induced by 
the  QCD-, electroweak- and magnetic penguins (the latter are induced by 
the dipole operators as will be discussed below).  
In general both QCD and EW penguins are important. In some cases, e.g. 
for the decays \rfn{ewd} for which QCD penguins are absent, the EW penguins 
dominate. To demonstrate the source of dominant new physics contribution
in LRSM we present as an example 
the Hamiltonian due to the gluon exchange
describing the decay $b\to q\bar q'q'$ at the scale $M_1$  
\bea
H^0_{eff}= -\frac{G_F}{\sqrt{2}}\frac{\alpha_s}{\pi} V_L^{tq*}V_L^{tb}
\left( \bar s\left[\Gamma_\mu^{LL}+\Gamma_\mu^{LR} \right] T^a b\right)
\left( \bar s \gamma^\mu T^a s \right),
\label{heffmw}
\eea
where the colour indices are understood and 
\bea
\Gamma_\mu^{LL}=E_0(x)\gamma_\mu P_L +
2i\frac{m_b}{q^2}E_0'(x)\sigma_{\mu\nu}q^\nu P_R, \nn \\
\Gamma_\mu^{LR}=2i\frac{m_b}{q^2}\tilde E_0'(x)
[A^{tb} \sigma_{\mu\nu}q^\nu P_R +
A^{tq*} \sigma_{\mu\nu}q^\nu P_L ].
\eea
Here the $\Gamma_\mu^{LR}$ term describes the new dominant
left-right contribution  which is induced by $W_1$ exchange which
interacts with right-handed currents via the mixing angle $\xi$
in one vertex of the penguin diagram and
\bea
A^{tb}=\xi \frac{m_t}{m_b} \frac{V_R^{tb}}{V_L^{tb}}e^{i\omega}\equiv
\xi \frac{m_t}{m_b} e^{i\sigma_1}, \nn \\
A^{tq}=\xi \frac{m_t}{m_b} \frac{V_R^{tq}}{V_L^{tq}} e^{i\omega}\equiv
\xi \frac{m_t}{m_b} e^{i\sigma_2}.
\label{a}
\eea
Note that the phases $\sigma_{1,2}$
are independent and can take any value in the range $(0,2\pi).$
 The functions $E_0(x),$ $E_0'(x)$ and
$\tilde E_0'(x)$ are Inami-Lim type functions \cite{lim} of 
$x=m_t^2/M_1^2$ and are given by
\bea 
E_0(x)&=&-\frac{2}{3}\ln x + \frac{x(18-11x-x^2)}{(12(1-x)^3)}+
\frac{x^2(15-16 x+4x^2)}{(6(1-x)^4)}\ln x, \nn \\
E_0'(x)&=& \frac{x(2+5x-x^2)}{(8(x-1)^3)}-
\frac{3x^2}{(4(x-1)^4)}\ln x, \nn \\
\tilde E_0'(x)&=& -\frac{(4+x+x^2)}{(4(x-1)^2)}+
\frac{3x}{(2(x-1)^3)}\ln x . 
\eea

It follows from \Eq{a}  that the new physics
contribution to the CP asymmetries in $B_s$ meson decays via \Eq{heffmw},
is suppressed by the bounds on the 
left-right mixing angle $\xi\lsim 0.01$ \cite{lang}, 
but  enhanced due to
\begin{enumerate}
\item[$ (i)$] the large ratio $m_t(M_Z)/m_b(M_Z)=60$ 
for $m_t(M_Z)=170$ GeV and $m_b(M_Z)=2.8$ GeV \cite{pich}. This enhancement 
factor arises due to the presence of $(V+A)$ interactions since no helicity 
flip in external $b$ quark line is needed in penguin contributions.
\item[$ (ii)$] the large value of the loop function $\tilde E_0'(x)$
which is is numerically about factor of four larger than the SM function
$E_0'(x)$.
\item[$ (iii)$] the {\it two independent} new phases $\sigma_{1,2}$ whose
values are unconstrained.
\end{enumerate} 
Note that $(i)$ together with $(ii)$ completely overcome the left-right 
suppression due to the smallness of $\xi.$ 
Therefore large effects are anticipated in 
CP asymmetries due to  $(iii).$

To calculate $B_s$ meson decay rates at the energy scale $\mu=m_b$ 
in the leading logarithm (LL) 
approximation we adopt the procedure and results 
from Ref. \cite{misiak}.
Using the operator product expansion to integrate out the heavy 
fields, and to calculate  the LL Wilson coefficients $C_i(\mu)$  
we run them with the renormalization 
group equations from the scale of $\mu=W_1$ 
down to the scale $\mu=m_b$ (since the contributions of $W_2$
are negligible we  start immediately from the $W_1$ scale).  
Because the new physics appears only in the  
magnetic dipole  operators we can safely take over some well-known
results from the SM studies. 
Therefore the the LRSM effective Hamiltonian  should 
include only these new terms which mix with the gluon and 
photon dipole operators under QCD renormalization. 
We work with the effective Hamiltonian 
\bea
H_{eff}&=&\frac{G_F}{\sqrt{2}} \left[
V_L^{uq*}V_L^{ub}
\sum_{i=1,2} C_i(\mu) O^u_i(\mu) +
V_L^{cq*}V_L^{cb}
\sum_{i=1,2} C_i(\mu) O^c_i(\mu) \right. \nn \\
&-& \left. 
V_L^{tq*}V_L^{tb}\left(
\sum_{i=3}^{12} C_i(\mu) O_i(\mu) +
 C^{\gamma}_7(\mu) O^{\gamma}_7(\mu) + C^{G}_7(\mu) O^{G}_7(\mu)
\right)\right] + (C_iO_i \to C'_iO'_i)\,,
\eea
where $O_{1,2}$ are the standard current-current operators,
$O_3$-$O_{6}$ and $O_7$-$O_{10}$ are the standard QCD and EW penguin
operators, respectively, and  $O_7^\gamma$ and $O_8^G$ are the
standard photonic and gluonic magnetic operators, respectively.
They can be found in the literature (e.g. Ref. \cite{ag,akl,cct}) and 
we do not present them here.
The new operators to be added, $O_{11,12},$ 
are analogous to the current-current
operators  $O_{1,2}$ but with different chiral structure  \cite{misiak}
\bea
O_{11}&=& \frac{m_b}{m_c}(\bar s_\alpha\gamma^{\mu}(1-\gamma_5) c_\beta )
(\bar c_\beta \gamma_\mu (1+\gamma_5)b_\alpha), \nn \\
O_{12}&=& \frac{m_b}{m_c}(\bar s_\alpha\gamma^{\mu}(1-\gamma_5)c_\alpha )
(\bar c_\beta \gamma_\mu (1+\gamma_5)b_\beta).
\eea
 Due to the left-right symmetry the operator basis is doubled by
including operators  $O'_i$ which can be obtained from $O_i$ 
by the replacements $P_L\leftrightarrow P_R.$  

Because the new physics affects only the Wilson coefficients
$C_7^\gamma,$ $C_8^G$ and $C_7^\gamma,$ $C_8^G$ it is sufficient 
to consider the basis $O_{1-6},O^\gamma_7,O^G_8,O_{11,12}$ + $(O\to O')$
for calculating them in the LL precision. 
Keeping only the top and bottom  quark masses
non-vanishing, the matching conditions 
at $W_1$ scale are given as 
\bea
C_2(M_1)=1,\qquad  \qquad  \qquad \qquad  \qquad  \qquad  
&&C'_2(M_1)=0, \nn \\
C^\gamma_7(M_1)=D_0'(x)+A^{tb}\tilde D_0'(x),   
\qquad\qquad  
&&C'^\gamma_7(M_1)=A^{ts*}\tilde D_0'(x), \nn \\  
C^G_8(M_1)=E_0'(x)+A^{tb}\tilde E_0'(x), 
\qquad \qquad  
&&C'^G_8(M_1)=A^{ts*}\tilde E_0'(x),
\eea
and the rest of the coefficients vanish.  
Here the SM function $D_0'(x)$  and 
its left-right analog $\tilde D_0'(x)$ are given by
\bea
D_0'(x)&=&\frac{ x(7-5x-8x^2)}{(24(x-1)^3)}-
\frac{x^2(2-3x)}{(4(x-1)^4)}\ln x, \nn \\
\tilde D_0'(x)&=& \frac{(-20+31x-5x^2)}{(12(x-1)^2)}+
\frac{x(2-3x)}{(2(x-1)^3)}\ln x\,. 
\eea
The $20\times 20$ anomalous dimension  matrix decomposes into two identical
$10\times 10$ sub-matrices. The SM $8\times 8$ 
sub-matrix of the latter one can be found in Ref. \cite{rome} and
the rest of the entries have been calculated by Cho and Misiak in Ref. 
\cite{misiak}.
In the LL approximation the low energy Wilson 
coefficients for five flavours are given by 
\bea
C_i(\mu=m_b)=\sum_{k,l}(S^{-1})_{ik}(\eta^{3\lambda_k/46})S_{kl} C_l(M_1),
\eea
where the $\lambda_k$'s in the exponent of $\eta=\alpha_s(M_1)/\alpha_s(m_b)$
are the eigenvalues of the anomalous dimension matrix over $g^2/16\pi^2$
and the matrix $S$ contains the corresponding eigenvectors. 
One finds for photonic magnetic coefficients \cite{misiak}
\bea
C^\gamma_7(m_b)&=&C_7(m_b)_{SM}+ 
A^{tb}\left[ \eta^{\frac{16}{23}}\tilde D_0'(x)+\frac{8}{3}
\left( \eta^{\frac{14}{23}}-\eta^{\frac{16}{23}} \right)
\tilde E_0'(x)\right]\,, \nn \\
C'^\gamma_7(m_b)&=&
A^{tq*}\left[ \eta^{\frac{16}{23}}\tilde D_0'(x)+\frac{8}{3}
\left( \eta^{\frac{14}{23}}-\eta^{\frac{16}{23}} \right)
\tilde E_0'(x)\right]\,,
\eea
where 
\bea
C^\gamma_7(m_b)_{SM}=\eta^{\frac{16}{23}}D_0'(x)+\frac{8}{3}
\left( \eta^{\frac{14}{23}}-\eta^{\frac{16}{23}} \right)E_0'(x)+
\sum_{i=1}^8 h_i\eta^{p_i}\,,
\eea
where  $h_i=(2.2996,$ -1.0880, -0.4286, -0.0714, -0.6494, -0.0380, -0.0186, 
-0.0057) and $p_i=(0.6087,$ 0.6957, 0.2609, -0.5217, 0.4086, -0.4230, 
-0.8994, 0.1456).
Similarly one finds for the gluonic magnetic coefficients
\cite{meie}
\bea
C^G_8(m_b)&=&\eta^{\frac{14}{23}}(E_0'(x)+A^{tb}\tilde E_0'(x)) +
\sum_{i=1}^5 h'_i \eta^{p'_i}\,, \\
C'^G_8(m_b)&=&\eta^{\frac{14}{23}}A^{ts*}\tilde E_0'(x) \,,
\eea
where
$h'_{i}=(0.8623,$ -0.9135, 0.0209, 0.0873, -0.0571) and
$p'_{i}=(14/23,$ 0.4086, 0.1456, -0.4230, -0.8994).
Using $\Lambda^{(5)}_{\bar MS}=225$ MeV and $\mu=\bar m_b(m_b)=4.4$ GeV
the LL Wilson coefficients take numerical values:
\bea
&& C_1=1.144, \qquad\qquad C_2=-0.308,\qquad\qquad C_3=0.014, \nn \\
&& C_4=-0.030, \qquad\qquad C_5=0.009, \qquad\qquad C_6=-0.038, \nn \\
&& C_7=0.045\alpha, \qquad\quad\;\; C_8=0.048\alpha, \qquad\quad\;
C_9=-1.280\alpha, \nn \\
&& C_{10}=0.328\alpha, \qquad\qquad 
 C'^\gamma_7=-0.523 A^{tq*}, \quad\;
C'^G_8=-0.231 A^{tq*}, \nn \\
&& \;\;\;\;C^\gamma_7=-0.331-0.523 A^{tb}, \qquad 
C^G_8=-0.156-0.231 A^{tb}. 
\label{coef}
\eea

To calculate hadronic matrix elements of various $B_s$ decay modes
we use the factorization 
approximation which has been recently extensively discussed in the
literature \cite{bsw,ag,akl,cct}. 
Therefore we do not discuss it here but refer the reader
to the original literature. However, we have to explain the assumptions 
which are involved in evaluating the hadronic matrix elements
\bea
 \langle O^G_8\rangle &=&-\frac{2\alpha_s}{\pi}\frac{m_b}{q^2}
\langle (\bar q_\alpha i\sigma_{\mu\nu}q^\mu P_R T^a_{\alpha\beta} b_\beta )
(\bar q'_\gamma \gamma^\nu T^a_{\gamma\delta} q'_\delta)\rangle \,,
\label{o8}
\eea
 and similarly for $ \langle O'^G_8\rangle .$ 
Here $q^\mu$ is the momentum transfered by the gluon to the $(\bar q',q')$ 
pair.  We are interested in  two body decays of $B_s$ mesons. 
In the factorization approach the two quarks $\bar q'$ and $q'$ cannot
go into the same decay product meson due to color.
Following Ali and Greub \cite{ag} we therefore assume that the three momenta 
of  $\bar q'$ and $q'$ are equal in magnitude but opposite in direction, and
in this case one may assume
\bea
q^\mu=\sqrt{\langle q^2\rangle}\frac{p_b^\mu}{m_b}\,,
\label{q2}
\eea
where $\langle q^2\rangle$ is an averaged value of $q^2.$
Thus the parameter $\langle q^2\rangle$ introduces certain uncertainty
into the calculation. In the literature its value is varied in the range
$1/4 \lsim \langle q^2\rangle/m_b^2 \lsim 1/2$ \cite{hou}. 
In our numerical examples we use in the following 
$\langle q^2\rangle=1/2m_b^2.$

Combining \Eq{q2} with  \Eq{o8} and using the 
equation of motion and some colour algebra
relations  one easily finds that 
\bea
 \langle O^G_8\rangle &=&-\frac{\alpha_s}{4\pi}
\frac{m_b}{\sqrt{\langle q^2\rangle}}
\left[\langle O_4 \rangle+\langle O_6 \rangle - \frac{1}{N_c}
\left( \langle O_3 \rangle+\langle O_5 \rangle\right) \right] \,,
\eea
 and similarly for $ \langle O'^G_8\rangle .$ 
An identical procedure gives for $ \langle O^\gamma_7\rangle $ 
\bea
 \langle O^\gamma_7\rangle &=&-\frac{\alpha}{3\pi}
\frac{m_b}{\sqrt{\langle q^2\rangle}}
\left[\langle O_7 \rangle+\langle O_9 \rangle \right] \,,
\eea
 and similarly for $ \langle O'^\gamma_7\rangle .$

In the factorization approximation one can easily relate the hadronic
matrix elements of the operators $O_i$ and $O'_i.$ It is straightforward
to show that for the decays of the types 
$B_s\to PP,\;VV$ where $P$ and $V$ denote
any pseudo-scalar and vector meson, respectively, one has 
  $\langle O_i\rangle=-\langle O'_i\rangle$ while for the decays of the 
type $B_s\to PV$ one has $\langle O_i\rangle=\langle O'_i\rangle.$
Therefore the magnetic penguin contributions can be absorbed 
into penguin contributions by redefinitions of the Wilson coefficients
\bea
&& C_3^{eff}=C_3+\frac{1}{N_c}\frac{\alpha_s}{4\pi}
\frac{m_b}{\sqrt{\langle q^2\rangle}} \left( C^G_8+ n C'^G_8\right)\,,
\qquad\quad 
C_4^{eff}=C_4-\frac{\alpha_s}{4\pi}
\frac{m_b}{\sqrt{\langle q^2\rangle}} \left( C^G_8+ n C'^G_8\right)\,, \nn \\
&& C_5^{eff}=C_5+\frac{1}{N_c}\frac{\alpha_s}{4\pi}
\frac{m_b}{\sqrt{\langle q^2\rangle}} \left( C^G_8+ n C'^G_8\right)\,,
 \qquad\quad 
C_6^{eff}=C_6-\frac{\alpha_s}{4\pi}
\frac{m_b}{\sqrt{\langle q^2\rangle}} \left( C^G_8+ n C'^G_8\right)\,, \nn \\
&& C_7^{eff}=C_7- \frac{\alpha}{3\pi}
\frac{m_b}{\sqrt{\langle q^2\rangle}} 
\left( C^\gamma_7+ n C'^\gamma_7\right)\,,
\qquad\qquad \;
C_9^{eff}=C_9- \frac{\alpha}{3\pi}
\frac{m_b}{\sqrt{\langle q^2\rangle}} 
\left( C^\gamma_7+ n C'^\gamma_7\right)\,,
\label{ceff}
\eea
and for the remaining coefficients $C_i^{eff}=C_i.$
Here $n=+1$ for the  decays $B_s\to PV$, 
$n=-1$ for the  decays $B_s\to PP,\;VV$ and $N_c=3.$

Now we are ready to estimate the new physics contribution
to the $B_s$ decay amplitudes. We first study the pure penguin induced 
decay $B_s\to\phi\phi$ ($b\to s\bar s s$) which is dominated
by QCD penguins and  receives also 
about 30\% contribution from the EW penguins.
The branching ratio of this decay mode is large, of the order 
$B(B_s\to\phi\phi)\sim {\cal O}(10^{-5})$ \cite{cct}
which ensures detectability. The pollution from the other SM
diagrams is estimated to be below 1\% \cite{gross}. Since the CP
asymmetries in this mode should vanish in the SM  the decay 
$B_s\to\phi\phi$ should provide very sensitive tests of the SM.
The amplitude of this decay takes a form \cite{cct,fleish1}
\bea
A(B_s\to\phi\phi)=-\frac{G_F}{\sqrt{2}}V_L^{tb}V_L^{ts*}2 \left[
a_3+a_4+a_5-\frac{1}{2}(a_7+a_9+a_{10})\right]
X^{(B_s\phi,\phi)}\,,
\eea 
where $X^{(B_s\phi,\phi)}$ stands for the factorizable hadronic
matrix element and 
\bea
a_{2i-1}= C^{eff}_{2i-1}+\frac{1}{N_c}C^{eff}_{2i}\,,
\qquad\qquad
a_{2i}= C^{eff}_{2i}+\frac{1}{N_c}C^{eff}_{2i-1}\,.
\eea
The exact form of $X^{(B_s\phi,\phi)}$ can be found in Ref. \cite{cct}.
Since it cancels out in CP asymmetries we do not present it here.
Using $\sqrt{\langle q^2\rangle}=m_b/\sqrt{2},$ $\xi=0.01,$ $m_t/m_b=60$ and 
the numerical values of LL coefficients in \Eq{coef} we obtain
\bea
A(B_s\to\phi\phi)=-\frac{G_F}{\sqrt{2}}V_L^{tb}V_L^{ts*}
2 \left[-0.0164+0.0035\left(e^{i\sigma_1}-e^{-i\sigma_2}\right)\right]
X^{(B_s\phi,\phi)} \,.
\label{a1}
\eea 
This result implies large effects on CP asymmetries.
The maximum deviation from the SM prediction $a_{CP}=0$ is obtained if 
$\sigma_1=\pi-\sigma_2=\pi/2+\delta_D$
implying $(\bar A/A)_{max}=e^{0.85 i}$ and  $|a_{CP}^{(s) \,max}|=0.85.$

Next we consider the decays \rfn{ewd} which have the feature of having
only one isospin channel describing a $\Delta I=1$ transition. Consequently
QCD penguins cannot contribute to the processes. Since the current-current
operators contribution is CKM suppressed it has been shown \cite{cct} that 
the EW penguin contributions amount about 85\% of all the decay rates
which are estimated to be of order a few times  $10^{-7}.$ 
The amplitudes of the decays $B_s\to P,P$ and $B_s\to V,P$ (i.e., 
pseudoscalar factorized out) receive a dominant contribution
from the terms proportional to $(-a_7+a_9).$ In the light of 
\Eq{ceff} it is clear that in this case the new physics contribution 
cancels out. However, the decays of type $B_s\to P,V$ and $B_s\to V,V$
depend on $(a_7+a_9)$ and they should receive sizable new physics 
contributions. For definiteness
let us work with the process $B_s^0\to \eta\rho^0$ 
but the results apply also for the decays $B_s^0\to \eta'\rho^0,\,\phi\rho^0.$ 
The amplitude can be written as
\bea
A(B_s\to\eta\rho^0)&=&\frac{G_F}{\sqrt{2}} \left[
V_L^{ub}V_L^{us*} a_2 
-V_L^{tb}V_L^{ts*}\left( \frac{3}{2}(a_7+a_9)\right) 
\right] X^{(B_s\eta,\rho)}_u \nn \\
&=&\frac{G_F}{\sqrt{2}} |V^{ts}_L|\left[
\lambda^2R_be^{-i\gamma} a_2 
+\left( \frac{3}{2}(a_7+a_9)\right) 
\right] X^{(B_s\eta,\rho)}_u\,,
\eea
 where $\gamma$ is the CKM angle, $X^{(B_s\eta,\rho)}_u$ is the hadronic
matrix element and $R_b=1/\lambda |V_L^{ub}|/|V_L^{cb}|.$
Because of the appearance of the CKM angle $\gamma$ in the decay 
amplitude Fleischer has proposed to use this process for determining
its value. In the SM one has 
$A(B_s\to\eta\rho^0)=A_{CC}e^{-i\gamma}+A_{EW},$ where $A_{CC}$ and $A_{EW}$
denote the current-current and EW penguin contributions, respectively.
While $A_{CC}\ll A_{EW}$ the fact that $A_{EW}$ is real still allows
for clean determination of $\gamma$ in the SM \cite{fleischer2}. However,
in the presence of new physics one has in general
\bea
A(B_s\to\eta\rho^0)=A_{CC}e^{-i\gamma}+A_{EW} +A_{NP}e^{-i\phi}\,,
\eea
where $A_{NP}$ is the magnitude of new contribution and $\phi$ its phase.
In such a case the CP asymmetry in the decay $B_s^0\to\eta\rho^0 $
takes a form
\bea
a_{CP}^{(s)}=\frac{2(y+\cos\gamma+z\cos\phi)(\sin\gamma+z\sin\phi)}{
y^2+2y(\cos\gamma+z\cos\phi)+1+z^2+2z\cos(\gamma-\phi)}\,,
\eea
where $y=A_{EW}/A_{CC}$ and $z=A_{NP}/A_{CC}.$
The corresponding SM expression is obtained by setting $z=0.$
For large $y\gg 1$ the CP asymmetry approaches 
$a_{CP}\to 2(\sin\gamma+z\sin\phi)/y$ and in the presence of 
sizable new physics contribution, $z\sim {\cal O}(1),$ the CP
asymmetry does not measure $\gamma$ any more.

In the LRSM the new contribution enters via the electromagnetic
dipole operators. In the presence of QCD penguins this contribution
is negligible because it is suppressed by $\alpha=1/128.$ 
However, in the present case QCD penguin contribution is exactly
vanishing and $A_{NP}$ should be compared with the sub-leading $A_{CC}.$
Using $\lambda=0.22,$ $|V_L^{ub}|/|V_L^{cb}|=0.08$ \cite{al} and rest of the
input as before we obtain numerically
\bea
A(B_s\to\eta\rho^0)=\frac{G_F}{\sqrt{2}}|V^{ts}_L|
\left[ -0.012+0.0013 e^{-i\gamma}+0.0011
\left(e^{i\sigma_1}+e^{-i\sigma_2}\right)\right] X^{(B_s\eta,\rho)}_u\,.
\eea
Therefore $z$ may be as large as $z\sim 2$ in the LRSM and 
comparison of $ a_{CP}(B_s\to\eta\rho^0)(t)$ with other measurements
determining CKM angle $\gamma$ could reveal the presence of new physics.

At this point a remark regarding the branching ratios is in order.
According to our results, modifications as large as 85$\%$ in the
CP asymmetries can be expected in the decays where new physics
modifies decay amplitudes. However, this cannot 
be directly translated to the branching ratios as, in this case,
large cancellations take place.
Even allowing maximum effects in the CP asymmetries the branching
rations are  modified not more than $\simeq 20 \% .$

\section{Conclusions}

In this work we have analyzed the possible effects of new physics 
in CP asymmetries in two body decays in left-right models with
spontaneous CP violation. This model possesses the attractive feature
that, quite independently of phenomenological considerations,
all the CP violating quantities (when the spontaneous
CP violation is achieved, as in our case, with the 
minimal content of the Higgs sector) depend on a single phase, $\alpha$,
and not on unconstrained quantities such as Yukawa couplings or additional 
phases.
This makes the model particularly predictive.

We have shown that the width difference in left-right models 
can be drastically
modified, in fact, reduced from its Standard Model value. 
This reduction itself can be an indication
indication of CP violation. As the large Standard Model prediction
for $\Delta  \Gamma^{(s)}$ is a 
consequence that the decay width into CP even final
states is larger that that on to CP odd final states; 
the appearance of a new phase in the mixing amplitude can 
make the CP eigenstates very different from the mass eigenstates
and therefore both mass eigenstates can then be allowed to decay into 
CP even final states.

Considering the new contributions to the $B_s$ mixing, with its own phases
which in general differ from the SM one, as well 
as the new contribution to the penguin dominated decay amplitudes, 
we have found
that large deviations from the SM predictions are possible with the 
present constraints on the masses of new gauge and Higgs bosons and 
on the left-right mixing angle.
It is important to emphasize that these types of contributions
can be clearly differentiated by comparing the tree level 
 $W$-mediated diagrams (new physics enters here only in the
mixing) with the penguin dominated ones (both contributions
are present).

Due to the new physics contribution to
the $B_s$ mixing non-zero CP asymmetries as large as
$|a_{CP}|=1$ can appear even if
the SM predictions for them are negligible. Even more promising is,
perhaps, the fact that due to the new physics contributions to the decay
amplitudes their effect can be probed by comparing two experiments that
measure the same phase in the SM. 
CP asymmetries af all QCD penguin dominated 
decays may largely be affected by the new physics as we explicitly demonstrate 
in the case of $B_s^0\to\phi\phi.$
The EW penguin dominated decays 
$B_s^0\to \eta^{(')}\rho^0,\,\phi\rho^0$ may also receive sizable new 
contributions which dominate the CP asymmetry measurements.

\vspace{2.5cm}

\begin{center}
{\bf Acknowledgements}
\end{center}

We are very grateful to A. Ali and R. Fleischer  for 
enlightening discussions. 
G.B. acknowledges  the Graduiertenkolleg ``Elementarteilchenphysik bei 
mittleren und hohen Energien" of the University of Mainz for a 
post-doctoral fellowship.
J.M. acknowledges the financial support from a Marie Curie EC Grant
(TMR-ERBFMBICT 972147) and M.R. the grant from the Humboldt Foundation.
This work was supported in part by CICYT, under Grant AEN 96-1718.

\newpage

\begin{figure}[!ht]
  \begin{center}
  \epsfig{file=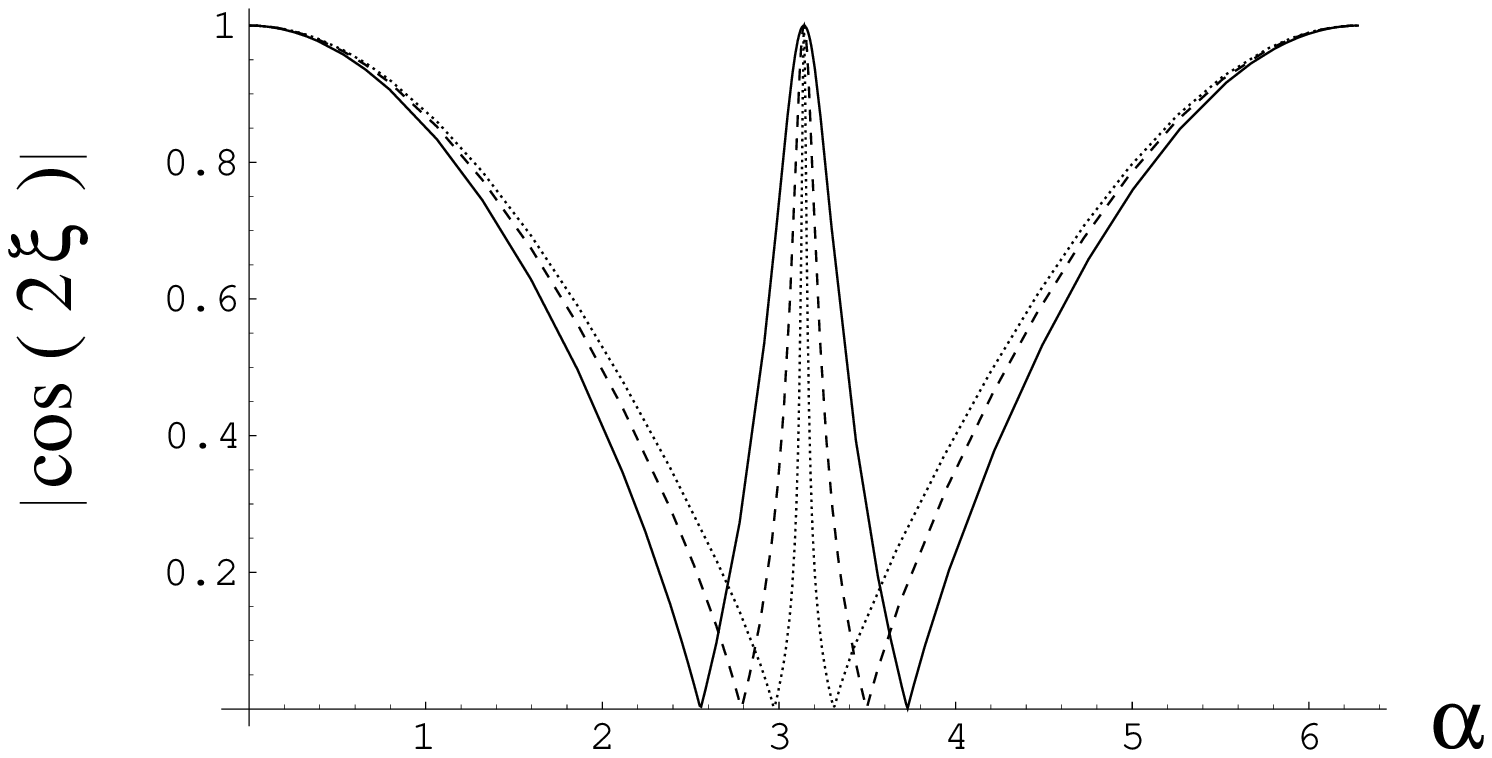,width=13cm}
%%%  \vspace{7cm} %%%
%%%  \parbox{15cm}{
\caption{Absolute value of $\cos(2\xi)$ (for the $B_s$ system) 
as a function of the
spontaneous symmetry breaking phase $\alpha$ for the fixed value
of the flavour changing Higgs mass $M_H=12$ TeV and for
the right handed gauge boson masses $M_2$ = 1.6 TeV (solid line),
5 TeV ( dashed line) and 9 TeV (dotted line).} %%% }
  \end{center}
\end{figure}
\vspace{1cm}

\begin{figure}[!ht]
  \begin{center}
  \epsfig{file=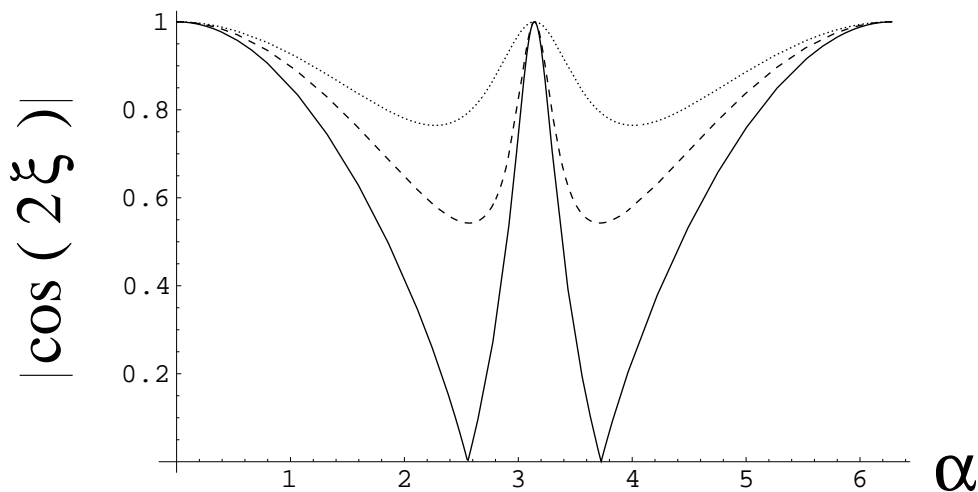,width=13cm}
%%%  \vspace{7cm} %%%
  \parbox{15cm}{\caption{Absolute value of $\cos(2\xi)$ 
(for the $B_s$ system) as a function of the
spontaneous symmetry breaking phase $\alpha$ for the fixed value
of the right-handed gauge boson mass $M_2=1.6$ TeV
and for the flavour changing Higgs boson masses $M_H$ = 12 TeV (solid line),
18 TeV ( dashed line) and 25 TeV (dotted line).}}
  \end{center}
\end{figure}

\end{document}